\documentclass[12pt]{article}
\usepackage{amssymb}
\usepackage{amsmath}
\usepackage{graphicx}
\usepackage{graphics}

\usepackage{color}

\title{How to recognize a nearly-flat-band ferromagnet by means of thermodynamic measurements?}

\author{Volodymyr Derzhko and Janusz Jedrzejewski \thanks{Corresponding author: jjed@ift.uni.wroc.pl}\\
Institute of Theoretical Physics, University of Wroc{\l}aw,\\
pl. Maksa Borna 9, 50--204 Wroc{\l}aw,
Poland}

\begin{document}
\maketitle

\begin{abstract}
We make an attempt at unveiling the thermodynamic ``signature'' of a specific class of electronic systems,
the so called nearly-flat-band paramagnets and ferromagnets that can theoretically be described by appropriate
versions of the Hubbard model.
\end{abstract}

\section{Introduction}

A nearly-flat-band system is an electronic system whose one or more bands in the single-particle spectrum
can be made arbitrarily narrow by continuously adjusting some parameters of the system, without destroying
its structure. To make this phrase more explicit, let us consider a tight-binding description of an electronic
system. In this description a single-particle system is given by a graph that consists of
a set of sites, with assigned values of external potentials, and a set of bonds connecting the sites,
associated to non-vanishing hopping intensities. Such a graph is {\em connected} if any two sites are
connected by a sequence of bonds.

Graphs of nearly-flat-band systems constitute a class of connected graphs, for which it is necessary to tune
the hopping intensities and on-site external potentials, or only one of those sets of parameters,
to get one or more degenerate energy levels, the so called {\em non-dispersive bands or flat bands},
whose degeneracy is proportional to the number of lattice sites.
As a result of the tuning, the hopping intensities and on-site external potentials satisfy some relations,
which constitute {\em sufficient conditions} for flat bands.
Note, however, that there are connected graphs (bipartite graphs) whose spectrum does contain a flat band,
and no tuning of hopping intensities is required \cite{Lieb-89}.

To each site of a graph there corresponds its {\em neighborhood}, that is the set of all the sites connected
with it by bonds. The {\em topology} of a graph is specified by the set of all the neighborhoods.
It is the topology, and not the geometry, of the underlying connected graph, that decides whether flat bands
can appear in a system. To the best of our knowledge, there is no characterization of the topology of graphs
that admit flat bands.
Only examples, sometimes of classes, of graphs where nearly flat bands do appear are known
\cite{Mielke-91/93,Tasaki-92,Mielke-Tasaki-93,OD-1,OD-2}.

By perturbing continuously the sufficient conditions for the flat band, that have to be
satisfied by hopping intensities and on-site external potentials, we transform a flat-band system into
a nearly-flat-band system.
This definition of a nearly-flat-band system, via a flat-band system is a theoretical, mathematical one.
The theoretical limit of non-dispersive band constitutes a nonphysical system that cannot be realized
in experiment.
The residual entropy of multi-electron free flat-band system, whose flat band is partially filled, is finite,
which violates the IIIrd law of thermodynamics, and there is no Fermi surface. The IIIrd law is violated
also in interacting systems; see \cite{OD-1, OD-2} for examples, where the ground-state degeneracy in an
interacting case was calculated exactly. Moreover, the density of states of those systems is singular.
In experiments, we can only attempt at constructing a nearly-flat-band system, where by suitably
adjusting the parameters of the underlying system we can make one or more bands arbitrarily narrow,
at least in principle.

Nevertheless, flat-band systems are worth of theoretical studies, since they are more simple for analysis
than nearly-flat-band systems, and moreover it might happen that some of their features are stable against
perturbations that transform a flat band into a nearly-flat band.
A prominent example of such a feature is ferromagnetism, and quite naturally the first model studied was the
paradigmatic Hubbard model:
\begin{eqnarray}
 H&=&\sum_{i,j,\sigma} t_{i,j}\,\,  c_{i,\sigma}^{+}c_{j,\sigma}
+ \frac {U}{2} \sum_{i,\sigma} n_{i,\sigma} n_{i,-\sigma},
\label{hubbard}
\end{eqnarray}
where the sums are over all the sites $i,j$ of the underlying graph, and over projections, $\sigma$,
of the electron spin on some axis; $t_{i,j}$ -- the matrix elements of a single-particle Hamiltonian
between states localized at sites $i$ and $j$ give the hopping intensities and on-site external potentials;
$c_{i,\sigma}^{+}$, $c_{i,\sigma}$, stand for the electron creation and annihilation operators, respectively;
the term proportional to $U>0$ represents a strongly screened  Coulomb repulsion.
When the underlying graph is bipartite, Lieb \cite{Lieb-89} proved the existence of {\em unsaturated ferromagnetism}
in the ground state (that is, for given number of electrons the total spin of the ground state is a fraction
of the maximal one), when it is a line graph -- a proof of {\em saturated ferromagnetism} in the ground state
(the total spin of the ground state is maximal) was given by  Mielke \cite {Mielke-91/93},
and when it belongs to Tasaki class -- the corresponding result was obtained by
Tasaki \cite{Tasaki-92, Mielke-Tasaki-93}.
In all those cases there is a flat band in the  single-electron spectrum, and on switching on the Hubbard repulsion
the paramagnetic ground state of a free multi-electron system turns into a ferromagnetic one, for a special value
of electron density  (or a narrow interval of densities) and  any nonzero value of the Hubbard on-site repulsion $U$.
The latter statement  means that paramagnetic ground state turns into a ferromagnetic one
without any competition between the kinetic and potential energies (the Hubbard repulsion is needed only to lift the
macroscopic degeneracy), which is another non-physical feature of flat-band systems.
The flat-band ferromagnetism discovered by Mielke and Tasaki appeared to be robust against perturbations of the flat band.
The proofs of this fact for nearly-flat band systems can be found in \cite{Tasaki-95/03,Tanaka-03}.

While there is a number of theoretical examples of nearly-flat-band systems (see the papers quoted above),
we do not know of any measurements performed on real nearly-flat-band systems.
There have been a few proposals of experimental realizations of nearly-flat-band systems such as:
atomic quantum wires \cite{Arita-98}, quantum-dot super-lattices \cite{Tamura-02}
or organic polymers \cite{Suwa-03}. However, the most promising seems to be a realization  as cold atoms in
optical lattices \cite{Bloch-08}. Due to a very good control of system parameters in the latter case,
such a realization would open new possibilities of investigating the mechanism of nearly-flat-band ferromagnetism,
inaccessible in other experimental realizations, and beyond the scope of present-day theoretical methods.
In this perspective, it is already interesting and challenging to determine theoretically characteristic
low-temperature thermodynamic properties of the aforementioned systems.

\section{The models and the goal}

In this paper, we provide a resume of our investigations of the question: how to recognize a nearly-flat-band ferromagnet
by means of thermodynamic measurements? In view of the high sensitivity of the considered systems to details of
the underlying graph, it would be naive to expect that some general answer, good for any type of such systems,
can be given.
We have chosen to concentrate on nearly-flat-band systems described by the Hubbard Hamiltonian,
with the lowest band nearly-flat and separated from the upper bands by a gap,
which we call the {\em principal gap}, to distinguish it from any other gap in the spectrum of the system.
Moreover, we set the Hubbard $U$ to be small compared to the value of the  principal gap, and the electron density
not to exceed $n_{fb}$, which is the density of those electrons that can be accommodated in the nearly-flat band.
For instance, the Tasaki models \cite{Tasaki-95/03} and some related models \cite{Ichimura-98} belong to this class.

One can argue, however, that our thermodynamic results, to be  presented below, hold as well for a wider class of systems.
This class includes systems where there are two groups of bands: low-energy bands and high-energy bands
with the nearly-flat band being the highest band in the group of the low-energy bands,
and separated by a principal gap from the upper bands \cite{Arita-98}.
Another group of systems that belongs to this class are the systems where the flat band sticks to a dispersive one
in a few points of the Brillouin zone, as in the Mielke class of models (see a remark in \cite{Kusakabe-94} and
numerical results in \cite{OD-2}).

To minimize the burden of large volume computer work we have performed calculations for the Hubbard model whose graph
is a  one-dimensional lattice, decorated with additional sites located in the middle between the  sites of the lattice
(known also as a $\Delta$-chain \cite{Ichimura-98} or a sawtooth chain \cite{OD-1}). For appropriate hopping intensities
and external potentials the lower band of the model is flat and the ground state is ferromagnetic for any $U>0$,
provided the electron density is $n_{fb}/2$. This model belongs to Tasaki class of models \cite{Mielke-Tasaki-93}.
There are many ways of perturbing the system to get a nearly-flat-band system, whose ground state is ferromagnetic
only for sufficiently large $U$. In our calculations we chose the Tasaki perturbation \cite{Tasaki-95/03},
whose advantage is that the ferromagnetic ground state and its energy are known explicitly.
The real matrix elements $t_{i,j}$ of the perturbed model can be chosen as follows
(with the lattice constant of the one-dimensional lattice set to unity):
$t_{i,i+1}=1$, $t_{i,i+1/2}=1+s$, $t_{i-1/2,i+1/2}=-s$, $t_{i,i}=2-s$, $t_{i+1/2,i+1/2}=1-2s$,
for an integer $i$, where $s >0$ is the parameter of Tasaki perturbation; up to a factor it amounts to the width of the
nearly-flat band. One finds that the width of the lower band is $\delta_{-}=4s$,
that of the upper band -- $\delta_{+}=4$, and the gap between the bands is $\varepsilon=1+s$.

For comparison, we consider also a fictitious noninteracting electron system, not born by a Hamiltonian,
whose single-particle spectrum consists of two bands, with some dispersion relations, separated by a gap.
The advantage of this model is  that the width of the lower band, $\delta_{-}$, the upper
band $\delta_{+}$, and the gap between the bands, $\varepsilon$, can be varied independently, what facilitates
observing their impact on thermodynamic properties of the model.

\section{The isochoric heat capacity}

We start our considerations of thermodynamic quantities with the entropy per particle, $s(T,v)$,
as a function of temperature, $T$ (throughout the paper $T$ is measured in energy units),
and volume per particle, $v$ -- the inverse of electron density, $n$.
Practically, since this entropy is not directly observable, we consider a simply related and accessible
to direct measurements quantity -- the isochoric heat capacity per particle (briefly specific heat), $c_V(T)$,
for specified values of electron density $n$.
The heat capacity at constant volume, is ideally suited for our purposes, since it describes a response of a system
to heating, exclusively due to thermal excitations. In contrast to other heat capacities, the isochoric heat capacity
contains no contribution of a mechanical or chemical work done by the system. Such an extra contributions blur the
response of the system to heating, making it more difficult to identify the system as a nearly-flat-band system.

Taking into account the variety of graphs and spaces of parameters of nearly-flat-band systems, we can claim that
ground states of a nearly-flat-band electronic systems are typically paramagnetic,
that is the total spin $S_{tot}= o(N_e)$, where $N_{e}$ is the number of electrons.
This is true even if we restrict the variety of graphs to those that admit ferromagnetic ground states,
that is states whose $S_{tot}= {\cal O}(N_e)$, for suitable values of system parameters.
The reason is that if the underlying graph admits ferromagnetism, rather limited values of electron densities,
nearly-flat band widths and Hubbard repulsion $U$ are required to make the ground state ferromagnetic.
Quite generally, there is no ferromagnetism, in the above sense, if the electron density
and/or Hubbard repulsion $U$ are too small, and/or the nearly-flat band -- too wide.

\begin{figure}[ht]
\centering
\includegraphics[width=0.70\textwidth]{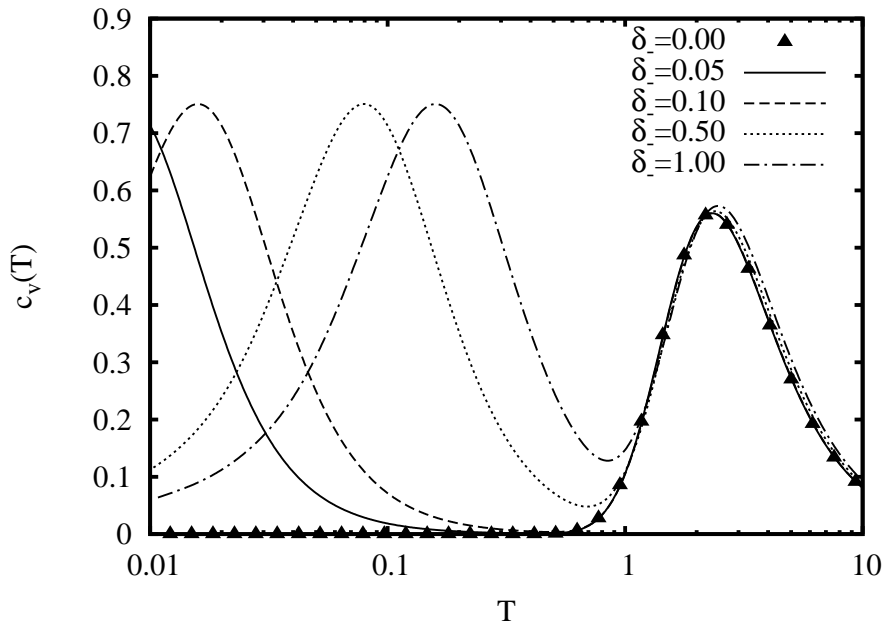}
\caption{Specific heat versus temperature, $c_V(T)$,
of the noninteracting two-band model, with $\delta_{+}=1$ and $\varepsilon=6$ (in  arbitrary energy units),
and partially filled lower band, for a decreasing sequence of lower-band widths $\delta_{-}$ and for the flat band.
Temperature is in a logarithmic scale.}
\label{cv2b}
\end{figure}

A principal gap in a single-particle spectrum induces a gap in the spectrum of the
corresponding many-electron noninteracting system and interacting system described by the Hubbard Hamiltonian.
Consequently, the spectra of these many-electron systems can be split into a lower part and an upper part.
As a result, plots of $c_V(T)$ consist of two, low-temperature and high-temperature, humps.
As long as our system is paramagnetic, the low-temperature hump is due to low-energy excitations, that is excitations
whose energies do not exceed the width of the lower part of energy levels  (the width of the nearly-flat
band in the noninteracting case). The high-temperature hump is, in turn, due to high-energy excitations, whose energy
does exceed the gap between the low-energy group  and the high-energy group  of energy levels (the principal-gap width
in a free nearly-flat-band system).
Consequently, the position of the low-temperature hump is  of the order of the width of the nearly-flat band,
while that of the high-temperature one is shifted by an energy of the order of the  principal gap,
at least for not too large $U$.
The degree of overlap of the two humps, i.e the extent and the depth of the  well between them, depends mainly on
the ratio of the principal gap and the width of the lower band.
The larger this ratio is the better the separation of the  humps. For sufficiently large gaps, the bottom of this well
reaches zero and is flat. Of course, $c_V(T)$ tends to zero if $T \to 0$ or $T \to \infty$.
These features are well illustrated in Fig.~\ref{cv2b}, where results for the two-band noninteracting system are shown.
In particular it is clear that this morphology of the $c_V(T)$ plot is no
characteristic of a nearly-flat-band system. It can be observed in any system of the class described above,
interacting or noninteracting, where instead of a nearly-flat band there is just a narrow band.

\begin{figure}[ht]
\centering
\includegraphics[width=0.70\textwidth]{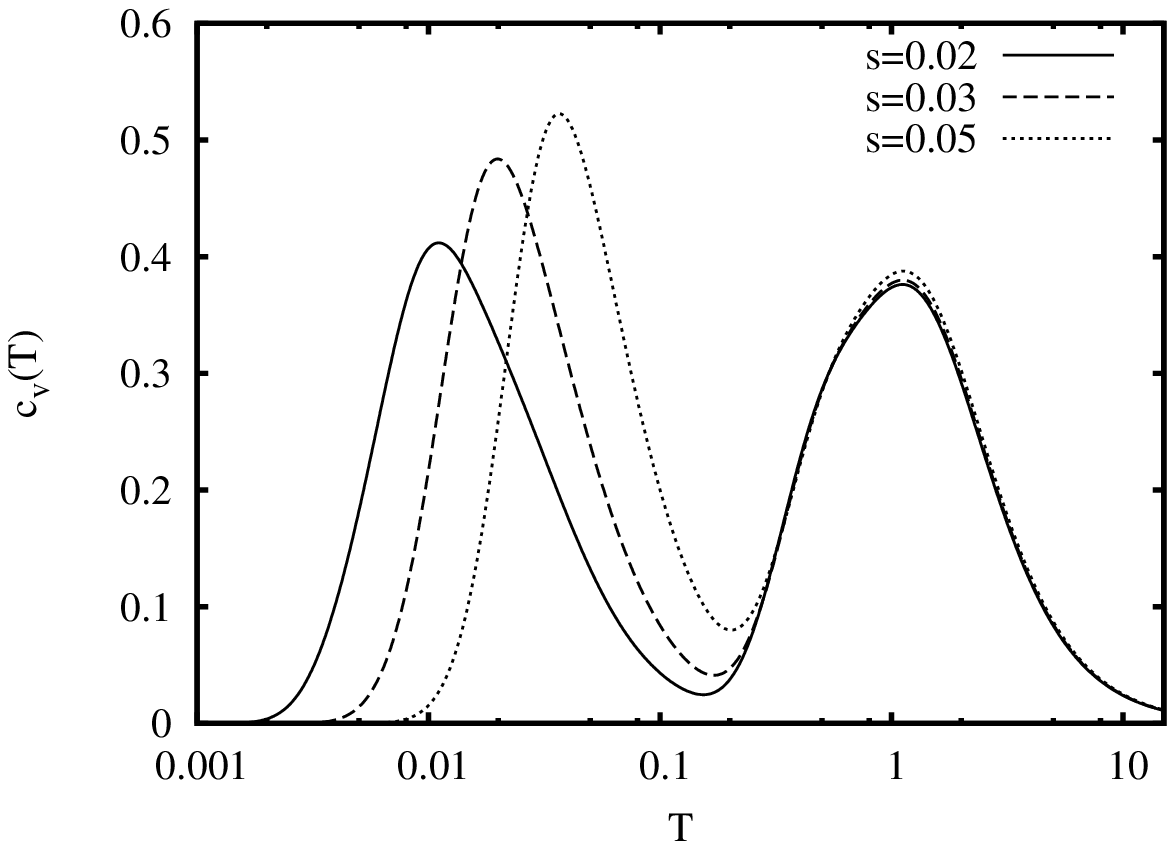}
\caption{Specific heat versus temperature, $c_V(T)$,
of the particular Hubbard model -- Tasaki model, defined in the paper, that consists of ten sites
with  periodic boundary conditions, and five electrons (quarter filling), for $U=0.1$ and for the range of
lower-band widths $\delta_{-}$, where a typical paramagnetic behavior is observed.
Temperature is in a logarithmic scale.}
\label{Hub1}
\end{figure}

In contrast to a narrow-band system, in a nearly-flat-band one we can exploit the possibility of shrinking
the lower band, while preserving the structure of our system.
On decreasing $\delta_{-}$, starting from sufficiently small value, the low-temperature hump moves, of course,
towards zero and almost linearly in $\delta_{-}$, its half-width shrinks, while its maximum remains
essentially unchanged. These effects can be seen in Fig.~\ref{cv2b} for the noninteracting two-band model and in Fig.~\ref{Hub1}
for the particular Hubbard model -- Tasaki model, defined in previous section.
In the theoretical limit of flat band, we are left with one, high-temperature hump, which may
be separated from zero temperature by a visible plateau of zero value, provided the principal gap is large enough.
This is because there is practically no low-temperature excitations from the flat band.

\begin{figure}[ht]
\centering
\includegraphics[width=0.70\textwidth]{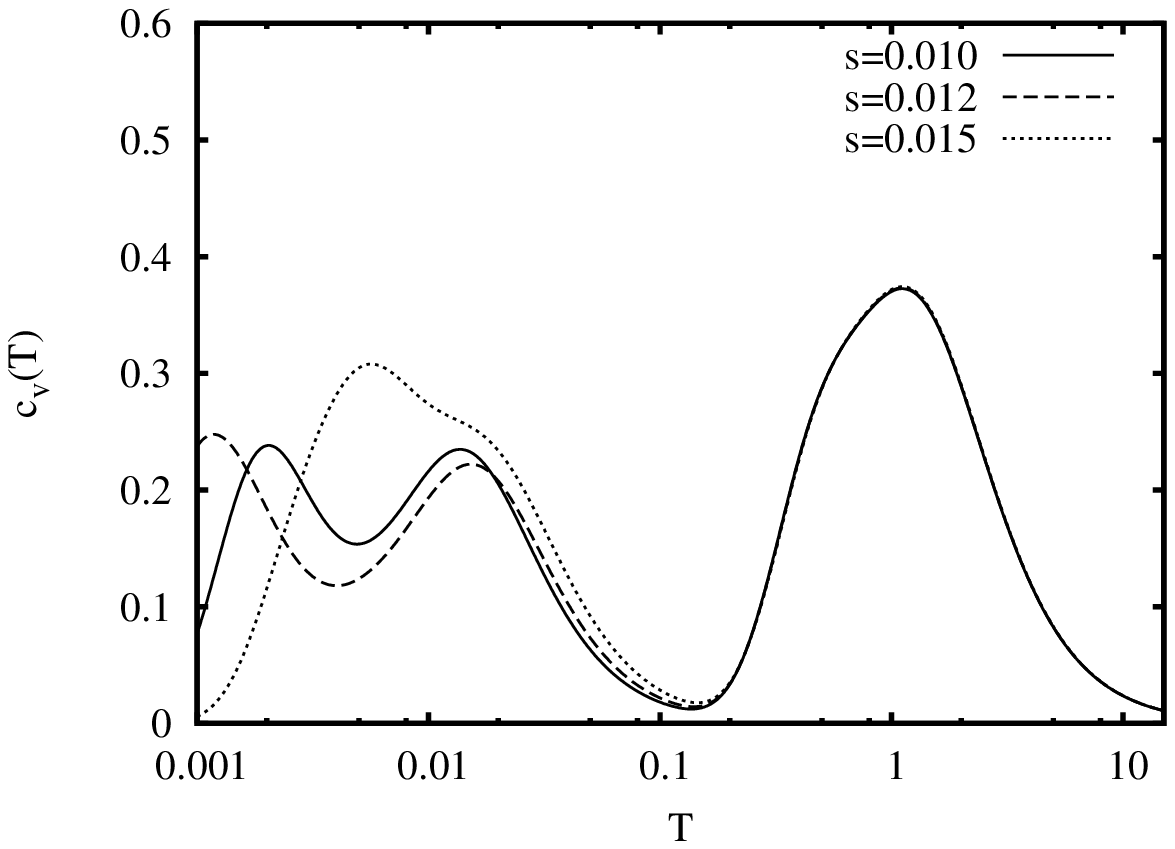}
\caption{Specific heat versus temperature, $c_V(T)$,
 of the particular Hubbard model -- Tasaki model, defined in the paper, that consists of ten sites
with  periodic boundary conditions, and five electrons (quarter filling), for $U=0.1$ and for the range of lower-band
widths $\delta_{-}$, greater than $\delta_{p-f}$ but smaller than the values of $\delta_{-}$ in Fig.~\ref{Hub1}.
Temperature is in a logarithmic scale.}
\label{Hub2}
\end{figure}

\begin{figure}[ht]
\centering
\includegraphics[width=0.70\textwidth]{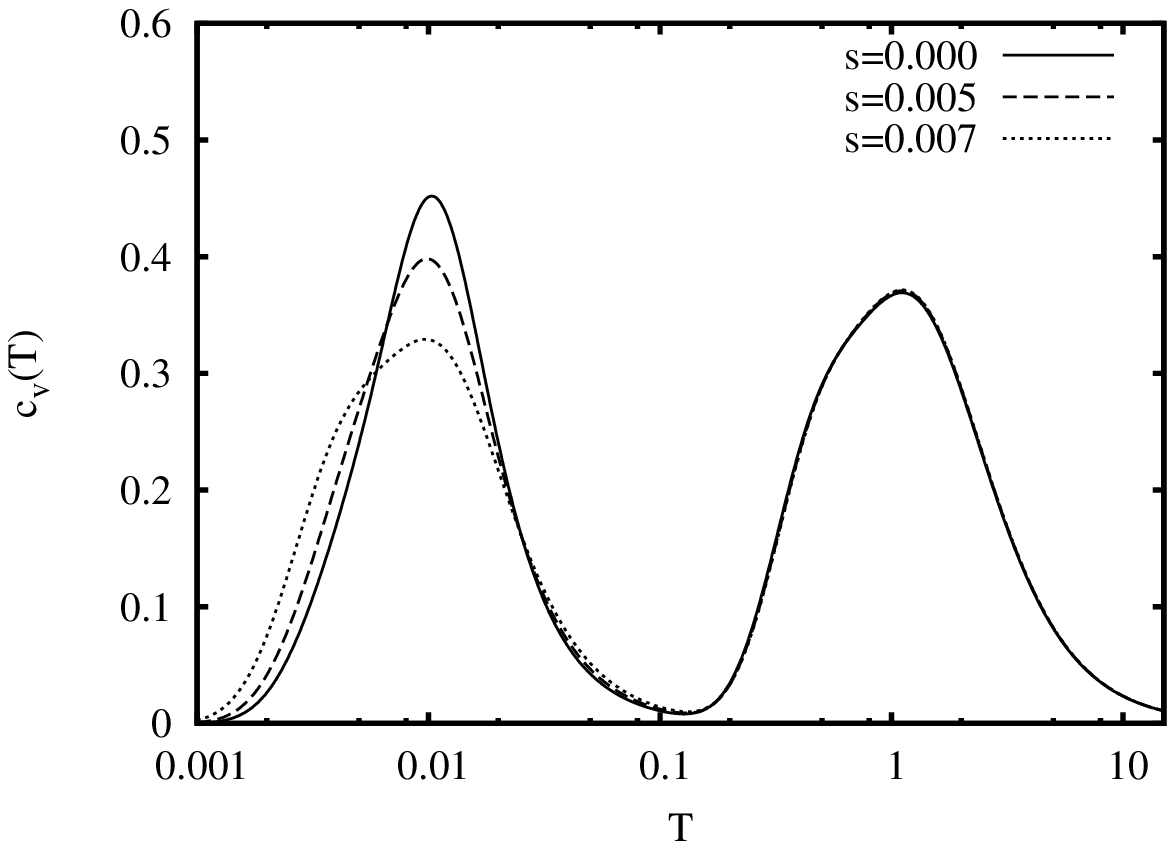}
\caption{Specific heat versus temperature, $c_V(T)$,
 of the particular Hubbard model -- Tasaki model, defined in the paper, that consists of ten sites
with  periodic boundary conditions, and five electrons (quarter filling), for $U=0.1$ and for the range of lower-band
widths $\delta_{-}$ that are smaller than $\delta_{p-f}$, and for the flat band.
Temperature is in a logarithmic scale.}
\label{Hub3}
\end{figure}

Now, suppose that the underlying graph, the electron density, and the Hubbard repulsion $U$ are such that
for those widths of the nearly-flat band that are smaller than some threshold value, $\delta_{p-f}$ (which depends on $U$),
the paramagnetic ground state changes into a ferromagnetic one.
Starting measurements with the values of $\delta_{-}$ somewhat larger than $\delta_{p-f}$, and then repeating them for
a decreasing sequence of values above $\delta_{p-f}$, one observes the above described ``evolution'' of the low-temperature
hump (see Fig.~\ref{Hub1}).
However, below $\delta_{p-f}$ the gapless low-temperature excitations of a paramagnet (of hole-particle type)
turn into ferromagnetic excitations -- the magnons.
This transition results in pinning the low-temperature hump at the temperature of the order of the width
of the magnons band, which is determined by $U$.
Further decrease of $\delta_{-}$, down to $\delta_{-}=0$, i.e. the limit of flat-band ferromagnet,
does not bring any significant changes to the low-temperature hump, neither to its position nor to its shape.
In Fig.~\ref{Hub2} we see a transient region between the paramagnet and the ferromagnet, with the values of $\delta_{-}$
greater than $\delta_{p-f}$, but smaller than the values of $\delta_{-}$ for which a typical for a paramagnet,
linear in $\delta_{-}$, ``motion'' of the low-temperature hump towards zero is observed.
In contradistinction to the overall shape of $c_V(T)$ plots in paramagnetic and ferromagnetic states, we expect that
the size (measured by a range of $\delta_{-}$) of the transient region and the shape of $c_V(T)$ plots in this region
are rather sensitive to the size of the system used for calculations.
Finally, in Fig.~\ref{Hub3},  we see the pinning of the low-temperature hump for $\delta_{-}$ smaller than $\delta_{p-f}$.

\section{Heat capacity at constant chemical potential}

In the previous section we discussed the heat capacity per particle at constant volume,
which is proportional to the second derivative with respect to temperature, $T$, of the fundamental thermodynamic function,
the Helmholtz free energy per particle as a function of $T$ and volume per particle, $v$.
An analog of this thermodynamic quantity, biased by finite-size effects, can be calculated numerically for small Hubbard
systems, with fixed volume and particle number, by means of the canonical ensemble.
Formally, the heat capacity at constant chemical potential, per unit volume, $c_{\mu}$, is another quantity of this kind.
It is proportional to the second derivative with respect to temperature of the fundamental thermodynamic function,
the grand-canonical potential per unit volume (i.e. minus the pressure) as a function of temperature and
chemical potential, $\mu$. This is, however, an unusual heat capacity, since it refers to open systems that are in
thermal equilibrium with particle reservoir of given chemical potential. Consequently, in a plot of $c_{\mu}$ versus $T$,
different points correspond, in general (depending on $\mu$, the electron density at constant $\mu$, $n_{\mu}(T)$,
can be monotonic or not), to systems with different amount of matter.
An analog of this thermodynamic quantity, biased by finite-size effects, can be calculated numerically for small Hubbard
systems, with fixed volume and chemical potential, by means of the grand-canonical ensemble.
Of course, $c_{\mu}(T)$ can be related to $n c_V(T)$, which is the isochoric heat capacity per unit volume,
by the thermodynamic identity:
\begin{equation}
c_{\mu}(T) = n_{\mu}(T) c_V(T,v(T,\mu)) +
T \left( \frac{\partial n_{\mu}(T)}{\partial T} \right)^{2}\left( \frac{\partial n_{\mu}(T)}{\partial \mu}  \right)^{-1}.
\label{cmu}
\end{equation}
The difference between those two heat capacities, $c_{\mu}-n_{\mu} c_V$, is born by the chemical work that an open system
does exchanging matter with matter reservoir when heated.

In a free system, whose spectrum consists of a flat band only, $c_V(T)$ vanishes identically, since there
are no excitations in this system. In contradistinction to $c_V(T)$, $c_{\mu}(T)$ can be nonzero, due to the chemical work,
the heated system does exchanging matter with a matter reservoir.
In such a system the electron density is a function of the activity $\xi = \exp(\mu/T)$ only;
we used here the fact that we can always set the energy of the flat band to zero.
Hence, the second term of identity (\ref{cmu}) -- the chemical work term can be written as
\begin{equation}
\xi \ln^{2}({\xi}) \, \frac{dn(\xi)}{d\xi} \geq 0.
\label{work}
\end{equation}
In a free system, with only a flat band, $dn(\xi)/d\xi > 0$, therefore the chemical work term, is nonzero unless
the chemical potential coincides with the energy of the flat band.
Consequently, there is just one hump in a plot of $c_{\mu}(T)$ for $\mu \neq 0$, located at low temperatures,
if $|\mu|$ is small. The size and the position of this hump is sensitive to the value of the chemical potential.

If a flat band is accompanied by some upper bands, separated by a gap from the flat one
(like in the systems considered in the previous section),
the above statements concerning low-temperature hump remain qualitatively true at sufficiently low temperatures
and for sufficiently small $|\mu|$. Additionally, such a system contributes significantly to $c_{\mu}(T)$
at high-temperatures; there is a high-temperature hump in the plot of $c_{\mu}(T)$, which coincides essentially
with $n c_V(T)$ -- the isochoric heat capacity per  unit volume. Small $|\mu|$ guarantees that the low- and
high-temperature humps are well separated.

\begin{figure}[ht]
\centering
\includegraphics[width=0.70\textwidth]{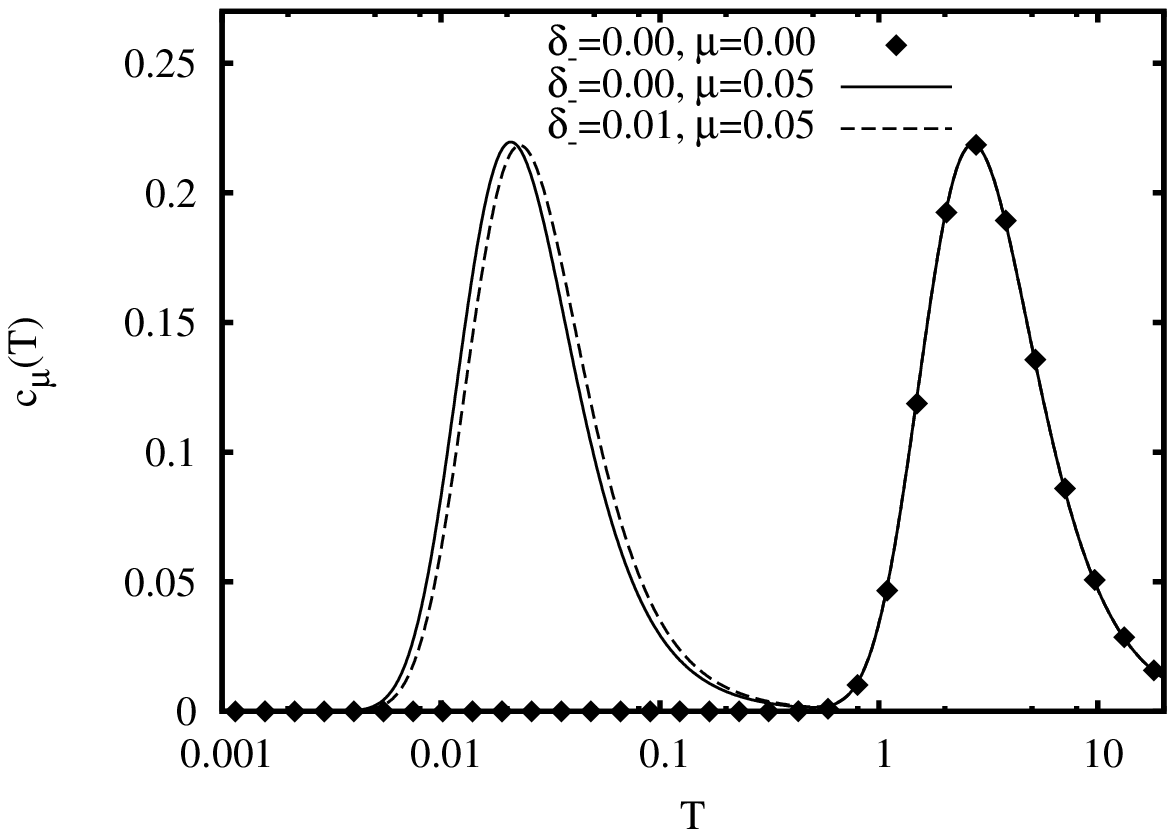}
\caption{Plot of $c_{\mu}(T)$ for the noninteracting two-band model, with $\delta_{+}=1$ and $\varepsilon=6$
(in  arbitrary energy units), with the flat-band energy and the upper edge of the corresponding nearly-flat band
set to zero. Temperature is in a logarithmic scale.}
\label{2bm_cm}
\end{figure}

\begin{figure}[ht]
\centering
\includegraphics[width=0.70\textwidth]{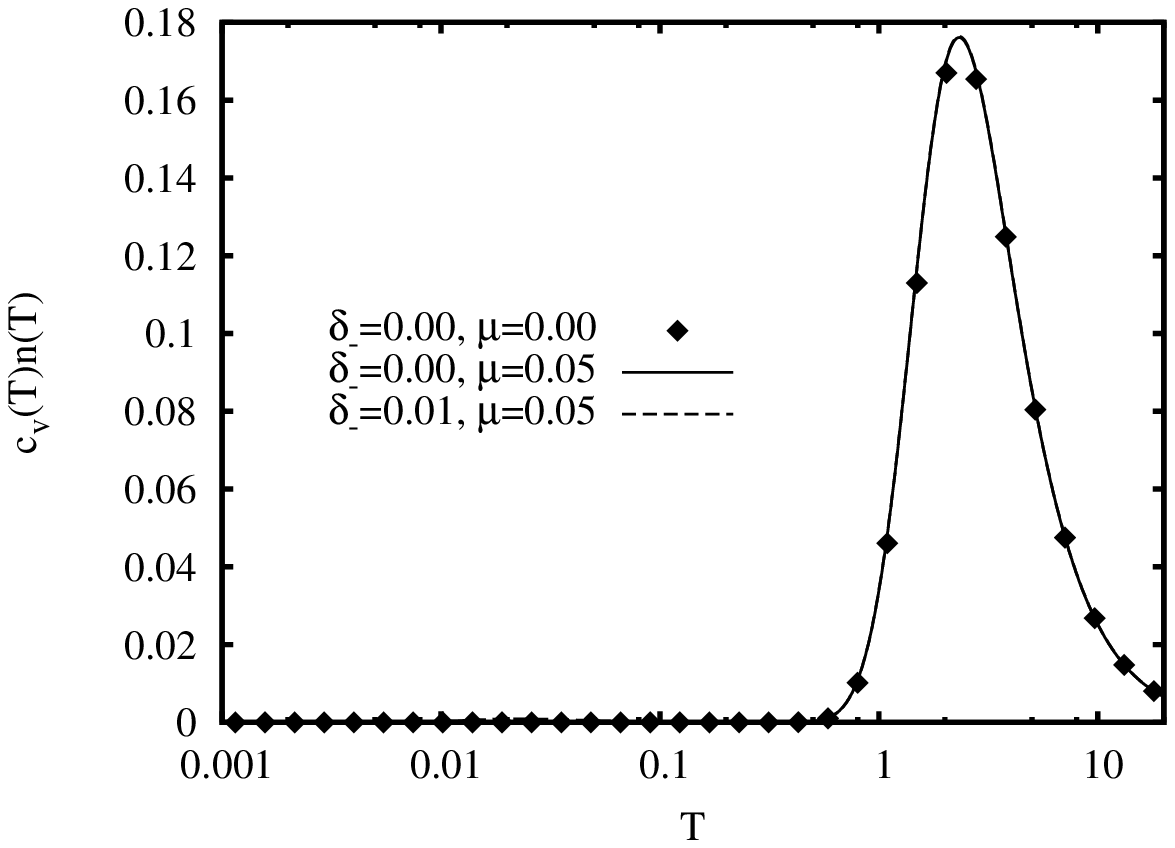}
\caption{Plot of $n_{\mu}(T)c_V(T,v(T,\mu))$ versus $T$, for the noninteracting two-band model, with $\delta_{+}=1$
and $\varepsilon=6$ (in  arbitrary energy units),
with the flat-band energy and the upper edge of the corresponding nearly-flat band set to zero.
The dashed line coincides with the continuous one.
Temperature is in a logarithmic scale.}
\label{2bm_cn}
\end{figure}

\begin{figure}[ht]
\centering
\includegraphics[width=0.70\textwidth]{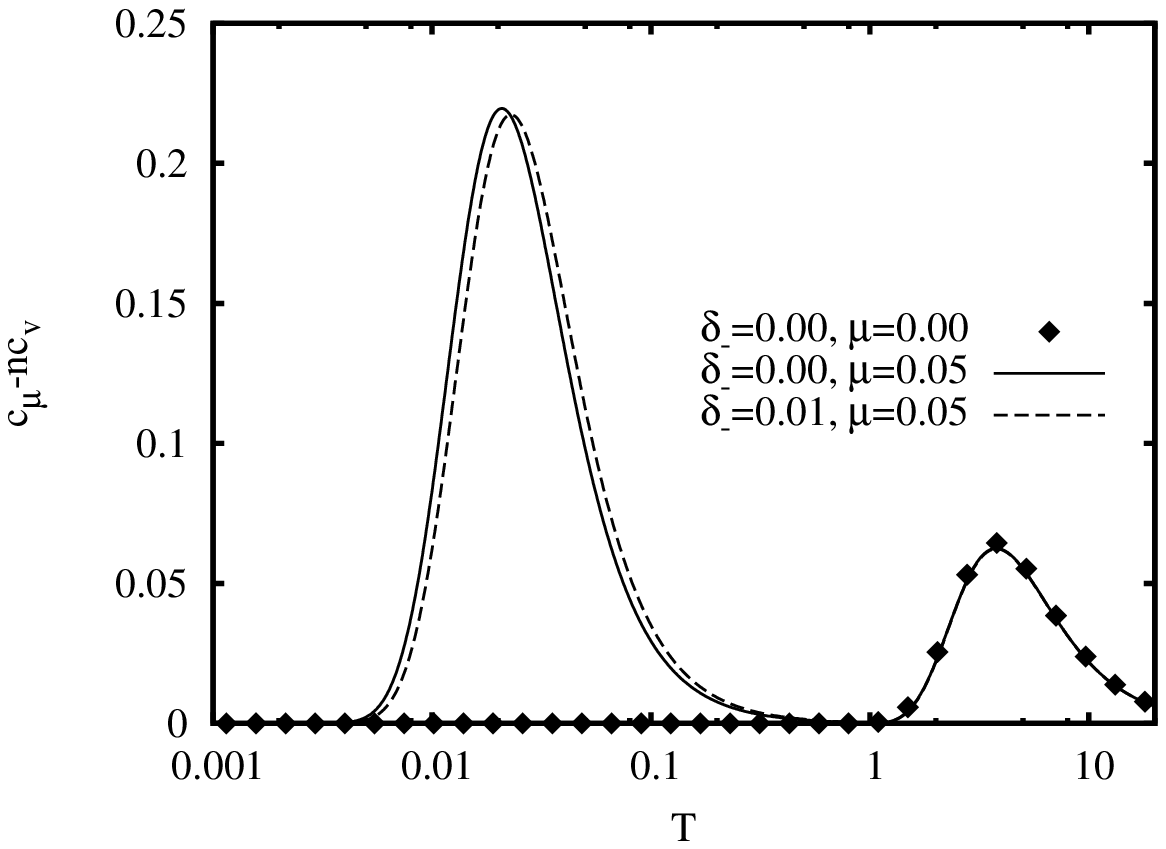}
\caption{Plot of $c_{\mu}(T)- n_{\mu}(T)c_V(T,v(T,\mu))$ versus $T$, for the noninteracting two-band model,
with $\delta_{+}=1$ and $\varepsilon=6$ (in  arbitrary energy units),
with the flat-band energy and the upper edge of the corresponding nearly-flat band set to zero.
Temperature is in a logarithmic scale.}
\label{2bm_re}
\end{figure}

A weak perturbation that turns a flat-band system into a nearly-flat-band one, could have influenced significantly
only the low-temperature hump.
Unlike the low-temperature hump in $c_V(T)$ plot of a free nearly-flat-band system, which is due to thermal
excitations from a nearly-flat-band, and therefore is sensitive to the width of this band (its position moves towards
zero temperature as the perturbation decreases), the low-temperature hump
in $c_{\mu}(T)$ is due to exchange of matter, and its position and shape are quite insensitive to the width of the
nearly-flat band. Thus, a weak perturbation of the flat band does not change essentially the two-hump $c_{\mu}(T)$ plot
of  the flat-band case.
The above observations are well illustrated in Figs.~\ref{2bm_cm},~\ref{2bm_cn},~\ref{2bm_re}.

Now, consider the Hubbard model (\ref{hubbard}) whose graph is specified in Section 2.
In the absence of perturbation ($s=0$), the lower band in the single-particle spectrum is flat
and is separated by a gap from the upper band.
There is a basis of the flat-band eigensubspace that consists of localized eigenstates.
In terms of Slater determinants of those localized single-particle eigenstates, one can construct
a basis of the ground-state eigensubspace of the Hubbard model, for any Hubbard repulsion $U>0$,
provided the number of electrons does not exceed half-filling of the flat-band \cite{Mielke-Tasaki-93}.
Derzhko et al \cite{OD-1} have demonstrated that those bases, as well as ground-state-eigensubspace bases of some
similar flat-band Hubbard models \cite{OD-2}, can be mapped onto a fictitious hard-core lattice gas.
The ground-state is paramagnetic, and the many-electron excited states are separated by a gap from the ground state,
for any filling that is smaller than half-filling of the flat-band.
Consequently, for such fillings, at sufficiently low temperatures that make excitations above the ground state
very unprobable, the response of the system to heating  amounts essentially to that of the hard-core gas corresponding
to the  macroscopically-degenerate ground state.
Since there are no excitations in a closed hard-core gas, $c_V$ vanishes identically. In contrast,
the chemical work term (\ref{work}), which in the case under consideration and in many other cases can be calculated
analytically, is nonzero for $\mu \neq 0$. Therefore, there is just one hump in a plot of $c_{\mu}(T)$ for $\mu \neq 0$,
at low temperatures.

A weak perturbation of the flat-band ($s \ll 1 $) does not bring essential changes to this low-temperature hump.
It is given essentially by the chemical work term of the hard-core gas
corresponding to the ground state of the flat-band limit of the considered nearly-flat-band Hubbard model.
Therefore, all the above described features of the low-temperature hump of $c_{\mu}(T)$ plot in free nearly-flat-band systems,
in particular its insensitivity to the width of the nearly-flat band, remain valid in the considered Hubbard model.
These observations apply also to other Hubbard systems mentioned in Section 2.
It is the derivative $dn(\xi)/d\xi$ that differentiates between hard-core gases and/or flat band systems
and makes the shape of the low-temperature hump specific for a system. However, those differences are not dramatic;
they are hidden in fine details of the shape of the low-temperature hump (see the plots in \cite{OD-2}).

Naturally, at high temperatures, there develops a high-temperature hump due to excitations above the ground state,
much like in the case of a free systems considered above, and it amounts essentially to the specific heat per  unit volume,
$nc_{V}(T)$.
The overall picture of $c_{\mu}(T)$ plot in the considered Hubbard models is qualitatively
the same as those in  free nearly-flat-band systems considered in the previous paragraph.
The described above features of $c_{\mu}(T)$ plot are well illustrated by the plots displayed
in \cite{OD-1}, for the Hubbard model considered in this paper, and in \cite{OD-2} for other Hubbard models with flat or
nearly-flat-bands. The data for all those plots were obtained from exact diagonalization of small Hubbard systems,
with the number of sites in their graphs between 10 and 20.
Apparently, looking at a $c_{\mu}(T)$ plot on can hardly infer whether the underlying system is a nearly-flat-band
(flat-band) Hubbard system.

\section{Summary}

Summing up, we propose an answer to the question asked in the title of our report. We argue that the isochoric
heat capacity per particle, $c_V(T)$, is a good candidate, a sufficiently sensitive thermmodynamic quantity, to measure.
At the heart of our answer is the fact that in the case of nearly-flat-band systems we can vary the width of the
nearly-flat lower band, and make it as narrow as we wish. Then, we can watch how  $c_V(T)$ plot ``evolves'' with
the decreasing width of the lower-band, at sufficiently low temperatures.
This low-temperature ``evolution'' provides a signature of a nearly-flat band paramagnet, paramagnet-ferromagnet transition,
and nearly-flat-band ferromagnet. A thermodynamic measurement performed only for a single value of a nearly-flat-band
width is not sufficient for this purpose.
We demonstrate also that the heat capacity at constant chemical potential, per unit volume, versus $T$, $c_{\mu}(T)$,
is not a suitable quantity to measure, since it depends weakly on the width of a nearly-flat-band and is sensitive to
the value of the chemical potential kept constant.
The shape of $c_{\mu}(T)$ plot is no characteristic of a nearly-flat-band (flat-band) Hubbard system.
One can ask naturally, whether such thermodynamic characteristics of closed systems like the coefficient of thermal
expansion or the isothermal compressibility can be used to recognize nearly-flat-band paramagnets and ferromagnets.
Our studies of the nearly-flat-band two-band model suggest that those quantities can be used to detect
nearly-flat-band Hubbard paramagnets. However, those quantities are not well defined and cannot be calculated for small
lattice systems. Therefore, we have no data for Hubbard systems with ferromagnetic ground state to check if those quantities
are suitable also for detecting nearly-flat-band ferromagnets.
\\[5mm]

\noindent
{\bf Acknowledgements}\\
\noindent
We thank Oleg Derzhko for discussions on flat-band systems.


\begin{thebibliography}{999}


\bibitem{Lieb-89}
E. H. Lieb,
{\em Two theorems on the Hubbard model},
Phys. Rev. Lett. {\bf 62}, 1201 (1989).

\bibitem{Mielke-91/93}
A. Mielke,
{\em Ferromagnetic ground states for the Hubbard model  on line graphs},
J. Phys. A: Math. Gen. {\bf 24}, L73 (1991).
{\em Ferromagnetism in the Hubbard model and Hund's rule},
Physics Letters A {\bf 174}, 443 (1993).

\bibitem{Tasaki-92}
H. Tasaki,
{\em Ferromagnetism in the Hubbard models with degenerate single-electron ground states},
Phys. Rev. Lett. {\bf 69}, 1608 (1992).

\bibitem{Mielke-Tasaki-93}
A. Mielke, H. Tasaki,
{\em Ferromagnetism in the Hubbard model},
Commun. Math. Phys. {\bf 158}, 341 (1993).

\bibitem{OD-1}
O. Derzhko, A. Honecker, and J. Richter,
{\em Low-temperature thermodynamics for flat-band ferromagnet: Rigorous versus numerical results},
Phys. Rev. B {\bf 76}, 220402(R) (2007).
A. Honecker, O. Derzhko, J. Richter,
{\em Ground-state degeneracy and low-temperature thermodynamics of correlated electrons on highly frustrated lattices},
Physica B {\bf 404}, 3316 (2009).


\bibitem{OD-2}
O. Derzhko, A. Honecker, and J. Richter,
{\em Exact low-temperature properties of a class of highly frustrated Hubbard models},
Phys. Rev. B {\bf 79}, 054403 (2009).
O. Derzhko, J. Richter, A. Honecker, M. Maksymenko, and R. Moessner,
{\em Low-temperature properties of the Hubbard model on higly frustrated one-dimensional lattices},
Phys. Rev. B {\bf 81}, 014421 (2010).


\bibitem{Tasaki-95/03}
H. Tasaki,
{\em Ferromagnetism in Hubbard models},
Phys. Rev. Lett. {\bf 75}, 4678 (1995).
{\em Ferromagnetism in the Hubbard model: a  constructive approach},
Commun. Math. Phys. {\bf 242}, 445 (2003).

\bibitem{Tanaka-03}
A. Tanaka and H. Ueda,
{\em Stability of ferromagnetism in the Hubbard model on the kagome lattice},
Phys. Rev. Lett. {\bf 90}, 067204-1 (2003).

\bibitem{Arita-98}
R. Arita, K. Kuroki, H. Aoki, A. Yajima, and M. Tsukada, S.
Watanabe, M. Ichimura, T. Onogi, and T. Hashizume,
{\em Ferromagnetism in a Hubbard model for an atomic quantum wire: a
realization of flat-band magnetism from even-membered rings},
Phys. Rev. B {\bf 57 }, R6854 (1998).

\bibitem{Tamura-02}
H. Tamura, K. Shiraishi, T. Kimura, and H. Takayanagi,
{\em Flat-band ferromagnetism in quantum dot superlattices},
Phys. Rev. B {\bf 65}, 085324 (2002).

\bibitem{Suwa-03}
Y. Suwa, R. Arita, K. Kuroki, and H. Aoki,
{\em Flat-band ferromagnetism in organic polymers designed by a computer simulation},
Phys. Rev. B {\bf 68}, 174419 (2003).

\bibitem{Bloch-08}
I. Bloch, J. Dalibard, W. Zwerger,
{\em Many-body physics with ultracold gases},
Rev. Mod. Phys. {\bf 80}, 885 (2008)

\bibitem{Ichimura-98}
M. Ichimura, K. Kusakabe, S. Watanabe, T. Onogi,
{\em Flat-band ferromagnetism in extended $\Delta$-chain Hubbard models},
Phys. Rev. B {\bf 58}, 9595 (1998).

\bibitem{Kusakabe-94}
K. Kusakabe and H. Aoki,
{\em Ferromagnetic spin-wave theory in the multiband Hubbard model having a flat band},
Phys. Rev. Lett. {\bf 72}, 144 (1994).

\end{thebibliography}
\end{document}